\global
\documentclass[pra,floatfix,nofootinbib, twocolumn,superscriptaddress,amsmath,amssymb,amsfonts]{revtex4-1} 

\usepackage[T1]{fontenc}
\usepackage{lmodern} % load a font with all the characters
\usepackage{color}
\usepackage{xcolor}
\usepackage{bbold}
\usepackage{dsfont}
\usepackage[draft,inline,nomargin]{fixme}
\usepackage{graphicx}
\usepackage[caption=false]{subfig}
\usepackage[colorlinks]{hyperref}
\usepackage[bottom]{footmisc}
\usepackage{enumerate}
\usepackage{float}
\usepackage{mathtools}
\usepackage{empheq}
\usepackage{tikz}
\usepackage{fancyhdr}
\usepackage[normalem]{ulem}
\DeclareMathAlphabet\mbc{OMS}{cmsy}{b}{n}
\usepackage{balance}

\begin{document}

\global\long\def\eqn#1{\begin{align}#1\end{align}}
\global\long\def\vec#1{\overrightarrow{#1}}
\global\long\def\ket#1{\left|#1\right\rangle }
\global\long\def\bra#1{\left\langle #1\right|}
\global\long\def\bkt#1{\left(#1\right)}
\global\long\def\sbkt#1{\left[#1\right]}
\global\long\def\cbkt#1{\left\{#1\right\}}
\global\long\def\abs#1{\left\vert#1\right\vert}
\global\long\def\cev#1{\overleftarrow{#1}}
\global\long\def\der#1#2{\frachttps://www.eluniversal.com.mx/techbit/search-2022-google-presenta-una-serie-de-novedades-en-sus-servicios
  {{d}#1}{{d}#2}}
\global\long\def\pard#1#2{\frac{{\partial}#1}{{\partial}#2}}
\global\long\def\re{\mathrm{Re}}
\global\long\def\im{\mathrm{Im}}
\global\long\def\dd{\mathrm{d}}
\global\long\def\ddd{\mathcal{D}}

\global\long\def\avg#1{\left\langle #1 \right\rangle}
\global\long\def\mr#1{\mathrm{#1}}
\global\long\def\mb#1{{\mathbf #1}}
\global\long\def\mc#1{\mathcal{#1}}
\global\long\def\tr{\mathrm{Tr}}
\global\long\def\dbar#1{\Bar{\Bar{#1}}}

\global\long\def\nth{$n^{\mathrm{th}}$\,}
\global\long\def\mth{$m^{\mathrm{th}}$\,}
\global\long\def\non{\nonumber}

\newcommand{\orange}[1]{{\color{orange} {#1}}}
\newcommand{\teal}[1]{{\color{teal} {#1}}}
\newcommand{\cyan}[1]{{\color{cyan} {#1}}}
\newcommand{\blue}[1]{{\color{blue} {#1}}}
\newcommand{\yellow}[1]{{\color{yellow} {#1}}}
\newcommand{\green}[1]{{\color{green} {#1}}}
\newcommand{\red}[1]{{\color{red} {#1}}}
\global\long\def\todo#1{\orange{{$\bigstar$ \cyan{\bf\sc #1}}$\bigstar$} }

\title{Effects of environment correlations on the onset of collective decay in
waveguide QED}

\author{A. Del \'{A}ngel}
\affiliation{Instituto de Investigaciones en Matem\'{a}ticas Aplicadas y en Sistemas, Universidad Nacional Aut\'{o}noma de M\'{e}xico, Ciudad Universitaria, 04510, DF, M\'{e}xico.}

\author{P. Solano}
\affiliation{Departamento de F\'isica, Facultad de Ciencias F\'isicas y Matem\'aticas, Universidad de Concepci\'on, Concepci\'on, Chile}
\affiliation{CIFAR Azrieli Global Scholars program, CIFAR, Toronto, Canada}

\author{P. Barberis-Blostein}
\affiliation{Instituto de Investigaciones en Matem\'{a}ticas Aplicadas y en Sistemas, Universidad Nacional Aut\'{o}noma de M\'{e}xico, Ciudad Universitaria, 04510, DF, M\'{e}xico.}

\begin{abstract}
  We calculate the dynamics of one and two two-level atoms interacting
  with the electromagnetic vacuum field in the vicinity of an optical nanofiber
  without making neither the Born nor the Markov approximations. We
  use a constant dielectric function and the Drude-Lorentz model,
  observing deviations from the standard super- and sub-radiant
  decays. Despite the non-trivial environment correlations, we discuss
  the validity of approximating the speed of atom-atom communication
  to the group velocity of the guided field. Our work presents a
  deeper understanding of the validity of commonly used approximations
  in recent platforms for quantum optics applications in the context
  of waveguide QED.
\end{abstract}

\maketitle

\section{Introduction}

The theory of open quantum systems studies the interaction of a
quantum system with its environment. Through a series of
approximations, one can describe the problem with simple equations,
which allow for analytical solutions in some particular cases
\cite{Carmichael2002,Carmichael1993book,rivas2012open,RevModPhys.89.015001,alicki2007quantum,BPbook}.
Perhaps the most relevant approximations assume that the quantum
system never entangles with its environment, known as the Born
approximation, and that the evolution time scale of the system is much
larger than the evolution of its environment, known as the Markov
approximation. Although these approximations accurately describe
common scenarios, such as atoms interacting through the
electromagnetic environment in free space
\cite{agarwal1974quantum,brooke_super-_2008}, one should question
their validity when describing novel experimental configurations
beyond what they originally intended to represent. In particular, the
rapidly growing field of waveguide quantum electrodynamics (wQED),
which describes atoms along a waveguide collectively interacting
through the guided field
\cite{Sheremet2021,Johnson2022,Han2021,Mirhosseini19,Wen19,Kim2018,Solano2017,Goban15,Asenjo17},
is built upon knowledge from open quantum systems. However, novel
experimental configurations \cite{Johnson2022,Han2021,Solano2017} and
theoretical proposals
\cite{Sinha2020,Sinha20,Solano2021,PhysRevLett.95.213001,PhysRevLett.110.113601,PhysRevResearch.2.013238,Olivera2022}, where the atoms interact with each other at large distances, raise
questions about the validity of the approximations borrowed from open
quantum systems interacting with the free space as their environment.

Quantitative analyses of the problem are intrinsically
system-dependent. Without losing sight of the phenomenology, we focus
our attention on optical nanofibers (ONFs) \cite{2017Solano}, a
platform that facilitates the interaction of emitters separated by
macroscopic distances. Ref. \cite{2005LeKien} presents the derivation
of a Markovian master equation for atoms interacting with an
electromagnetic environment in the presence of a nanofiber. The
authors assume that the distance between the atoms is negligible and
that the environment is Dirac delta-correlated. Nevertheless,
approximating the correlation functions of the electromagnetic
environment with a Dirac delta function is unrealistic since the field
emitted by an atom into the guided modes can strongly affect another
at a later time. A correlation function with two Dirac delta functions
separated by the delayed interaction time can simplify the problem,
interestingly leading to non-Markovian evolutions \cite{Sinha2020}.
Approximating the correlation functions of the electromagnetic
environment of a nanofiber with Dirac delta functions is so far a standard procedure in wQED; thus, proving its validity is crucial.
Besides estimating quantitative deviations from the predictions, a
detailed study of the correlation functions allows for answering
fundamental questions. For example, what is the field velocity that
accurately describes the delayed interaction time between two atoms:
phase or group velocity? what is the time scale and dynamics for the
appearance of collective behaviors? Is it possible to observe these
effects with current experimental technology?

In this paper, we study how the dynamics of two separated two-level atoms is affected by the correlations of the fundamental guided modes of an ONF at zero temperature, which
acts as the environment for the atoms. To do so, we calculate and
analyze in detail the correlation functions of the guided modes as a function of the separation between atoms. We consider two dispersion relations, a commonly assumed constant dielectric function and the more realistic Drude-Lorentz (DL) model for the nanofiber dielectric function. By numerically solving the dynamical equations,
we estimate the modification of the collective decay rates of the
atoms and explore the effects of explicitly considering the
correlations of the environment.

We show that for a single atom in the vicinity of an ONF at zero
temperature, it is unnecessary to modify the spectral density of the
environment to render its correlation close to a Dirac delta
distribution, contrary to the free space case
\cite{Carmichael2002,Rivas_2010}. In such a scenario, the Markovian
approximation is valid for the two dielectric functions we consider, a
result which, to our knowledge, has not been previously verified
despite its widespread use in this context. For two atoms, we observe
that the correlation functions for a constant dielectric function
resemble displaced Dirac delta distributions. However, the delayed
maxima of the environment correlations are not centered at nor
determined by the time it takes the electromagnetic field to propagate
between the atoms at the group or phase velocities. Nevertheless, when
we study the dynamics of the collective behavior, we obtain that the
onset of the collective behavior is consistent with assuming that the
atoms interact with a delay given by the group velocity only if the
atoms are distant enough. Additionally, we find that atoms prepared in
(anti)symmetric states radiate at rates slightly below (above) those
obtained with the Markovian approximation, suggesting the
impossibility of realizing perfect subradiant states. Our study
provides a test for the validity of usual approximations employed in
wQED.
%,such as the markovian \cite{2005LeKien} and the displaced Dirac delta correlations \cite{Sinha2020}.

%This may be attributed to the fact that the dispersion relation inside the ONF possesses the same asymptotic behaviour at low and high frequencies, regardless of the dielectric function used to model its response.

Our paper is organized as follows: in section
\ref{section:Physical_Model} we present the model for the system, and
describe how the correlation functions and the atoms' evolution are
numerically calculated. In section \ref{section:Results} we show and
discuss the main results of this work, as well as its implications.
Finally, we summarize and give an outline in section
\ref{section:Conclusions}.

\section{Physical Model}
\label{section:Physical_Model}

\begin{figure}[t]
\centering
     \includegraphics[width = 3 in]{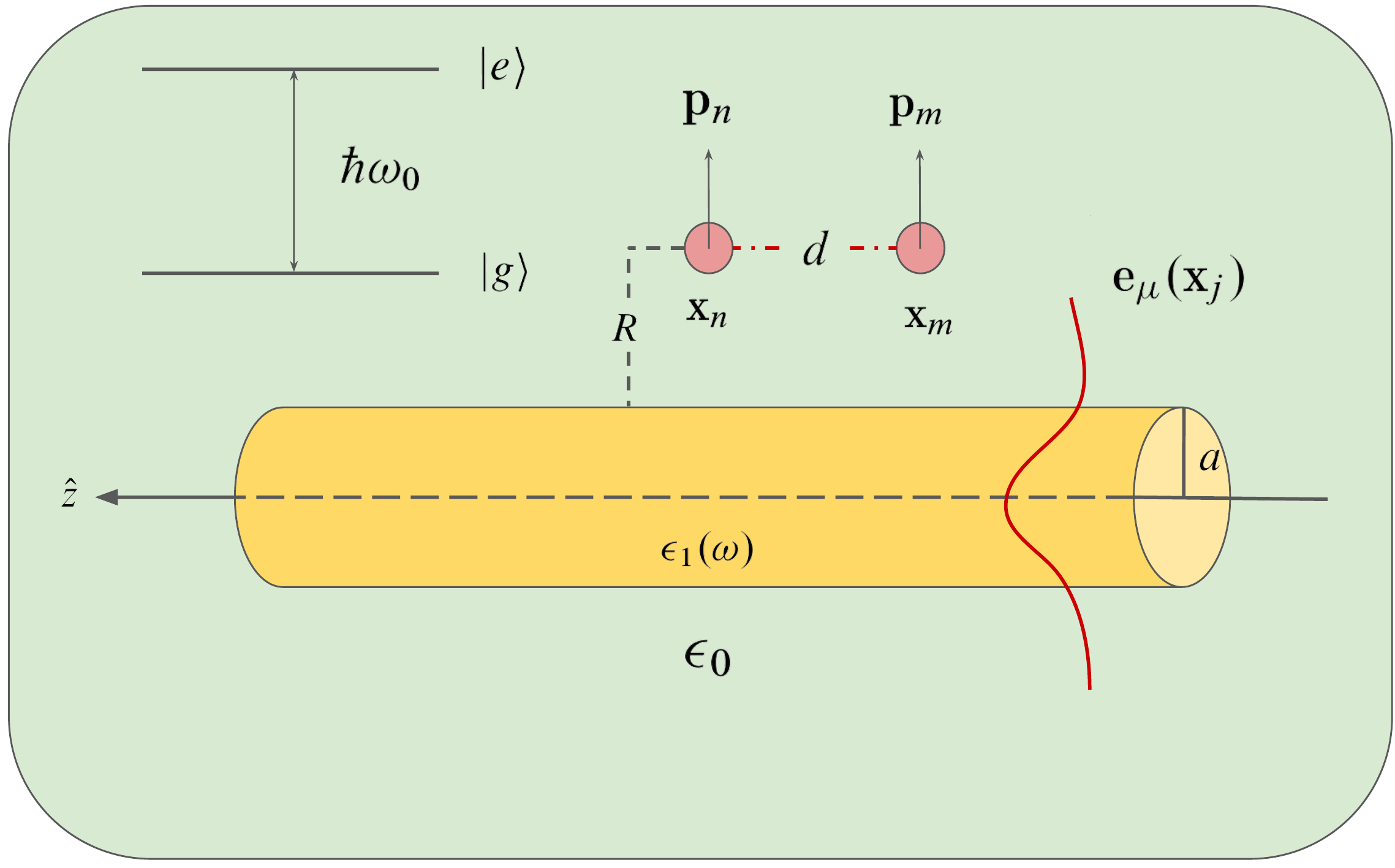}
    \caption{Schematic representation of the system. Two two-level atoms with resonant frequency $\omega_{0}$ separated a distance $d$ interact via a common bath, given by the evanescent field of the guided modes $\bf{e}_{\mu}$ of an optical nanofiber. The optical nanofiber of radius $a$ is characterized by a dielectric function  $\epsilon(\omega)$.}
    \label{Fig:sch}
\end{figure}

We consider two identical two-level atoms with resonance frequency
$\omega_{0}$ $(E_{e}-E_{g} = \hbar \omega_{0})$ in the vicinity of an optical nanofiber of radius $a$
and a frequency dependent dielectric function $\epsilon(\omega)$ (see Fig. \ref{Fig:sch}). We consider two dielectric
functions for our study: a commonly assumed constant function $\epsilon_{C}(\omega)\approx\epsilon_{C}(\omega_0)$, yielding a
constant refractive index
$n_{1} = \sqrt{\epsilon_{C}(\omega)/\epsilon_{0}} \approx 1.4534$, and
the more realistic Drude-Lorentz (DL) dielectric function
$\epsilon_{L}(\omega)/\epsilon_{0} = 1+
\omega_{p}^{2}[(\omega_{R}^{2}-\omega^{2})-i\gamma\omega]^{-1}$, where
$\omega_{R}$ and $\gamma$ are the resonant frequency and decay rate of the constituents of the ONF, respectively, and
$\omega_{p}$ is its plasma frequency \cite{Jackson}. The parameters of
the DL function are chosen so that they mimic the essential features
of silica glass, such as a high absorption for frequencies in the
ultraviolet regime and refractive indices close to 1.5 in the optical
regime \cite{2017Solano}. We take the values of these parameters to be
$\omega_{R}= \omega_{350} = 2\pi c/350\text{nm}$ and
$\gamma_{350} = 4\alpha a_{0}^{2}\omega_{350}^{3}/3c^{2}$,
corresponding to half of the ultraviolet interval and the decay rate
of a single constituent via an electric dipole transition,
respectively; the value of the plasma frequency is fixed by setting
the refractive index of the ONF to be $n_{1}$ at the resonance
frequency of the atoms. We neglect the effects of absorption
associated with the imaginary part of this function near the atomic resonance.

%Since we are not considering external fields other than the modes of the ONF, it is reasonable to assume that no higher order atomic transitions ought to occur and thus,
We study the case in which the atoms couple exclusively with the fundamental mode HE$_{11}$ of the guided field of the ONF by means of electric dipole interactions. In the interaction picture, the atoms-field interaction Hamiltonian after the rotating wave approximation (RWA) is given by
\begin{align}
    H_{\text{int}}  &= i\hbar\sum_{\mu}\sum_{m=1}^{2}G_{\mu m}e^{-i(\omega-\omega_{0})t}\sigma^{\dagger}_{m}a_{\mu} + \text{H.C.}\; , \label{Hamiltonian}\\
    G_{\mu m} &  = \sqrt{\frac{\omega\beta'}{4\pi\epsilon_{0}\hbar}} \bar{p}_{m}\cdot\bar{e}_{\mu}(\bar{x}_{m})e^{i(f\beta(\omega)z_{m}+\phi_{m})}.
\end{align}
Here, the index $m = 1,2$ labels an atom in position $\bar{x}_{m} = (r_{m},\phi_{m},z_{m})$ in cylindrical coordinates and the sum  $\sum_{\mu} = \sum_{l,f}\int_{0} ^{\infty}d\omega$ goes over the field polarization in the circular basis, the propagation direction along the fiber's axis and the frequency of the guided modes, which are encoded through the variable $\mu = (l = \pm 1,f = \pm 1 ,\omega)$; $\sigma_{m} = \ket{g_{m}}\bra{e_{m}}$ is the atomic lowering operator and $a_{\mu}$ is the annihilation operator of a photon with parameters $\mu$. The coupling frequencies $G_{\mu m}$ are written in terms of the propagation constant $\beta(\omega)$, the density of states $\beta' = \partial\beta/\partial\omega$, the
electric dipole matrix element of the m-th atom $\bar{p}_{m}$ and the components of the guided field modes $\bar{e}_{\mu}$, which are explicitly given in reference \cite{2005LeKien}. For each frequency component of the field, the propagation constant of the fundamental mode is obtained by numerically solving the following eigenvalue equation \cite{Marcuse} \small
\begin{align}
\frac{J_{0}(ha)}{haJ_{1}(ha)} & = \Big[\frac{n_{1}^{2}+n_{2}^{2}}{4n_{1}^{2}}\Big]\frac{K_{1}'(qa)}{qaK_{1}(qa)}+\Big(\frac{1}{ha}\Big)^{2}+R(\omega,\beta), \label{EVE} \\
    R(\omega,\beta) & = \Big(\Big[\frac{n_{1}^{2}-n_{2}^{2}}{4n_{1}^{2}}\frac{K_{1}'(qa)}{qaK_{1}(qa)}\Big]^{2} \notag \\
   & + \Big[\frac{\beta}{k_{1}}\Big(\frac{1}{(ha)^{2}} + \frac{1}{(qa)^{2}}  \Big) \Big]^{2} \Big)^{\frac{1}{2}},
\end{align}
where $n_{j}$ represents the refractive indices of the ONF (j = 1) and the vacuum (j = 2),  $h = \sqrt{k_{1}^{2}-\beta^{2}}$, $q = \sqrt{\beta^{2}-k_{2}^{2}}$, $k_{j} = n_{j}(\omega) \omega/c$ and $J_{j}$, $K_{j}$ are the j-th order Bessel functions of the first kind and the modified of the second kind, respectively.\\

\begin{figure*}[t]
    \centering
    \subfloat[]{\includegraphics[width = 3 in]{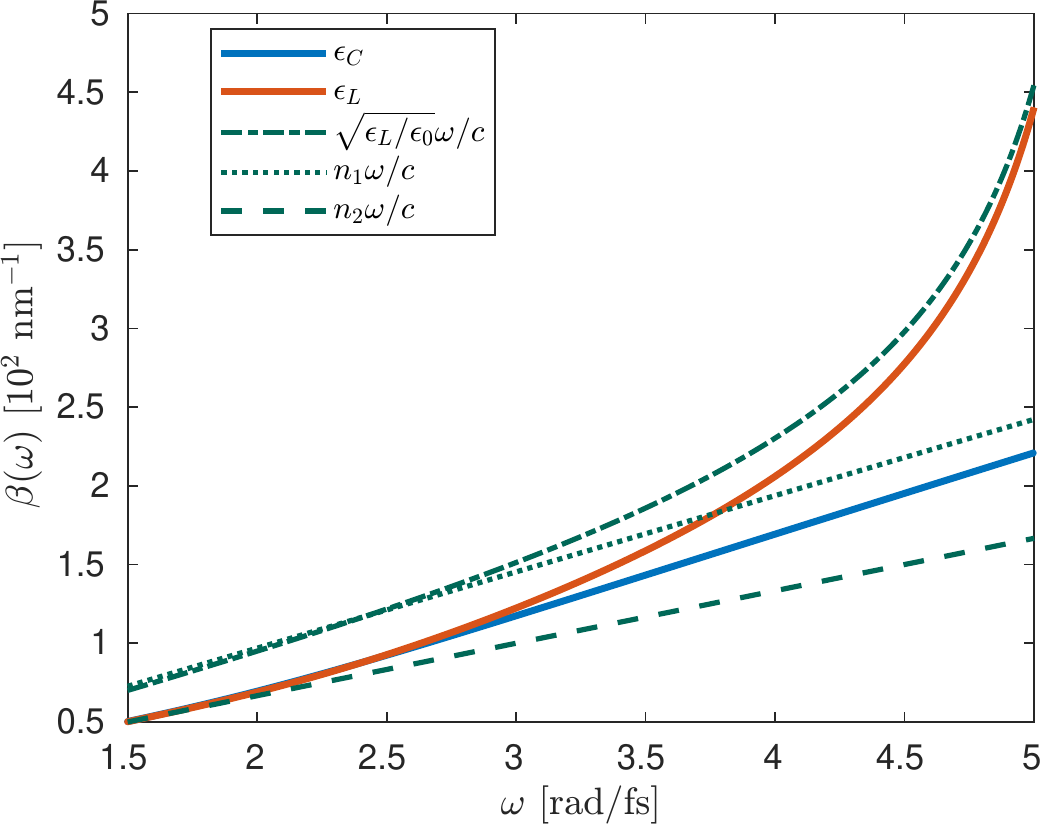}
    \label{Fig:DispRel}}
    \subfloat[]{\includegraphics[width = 3 in]{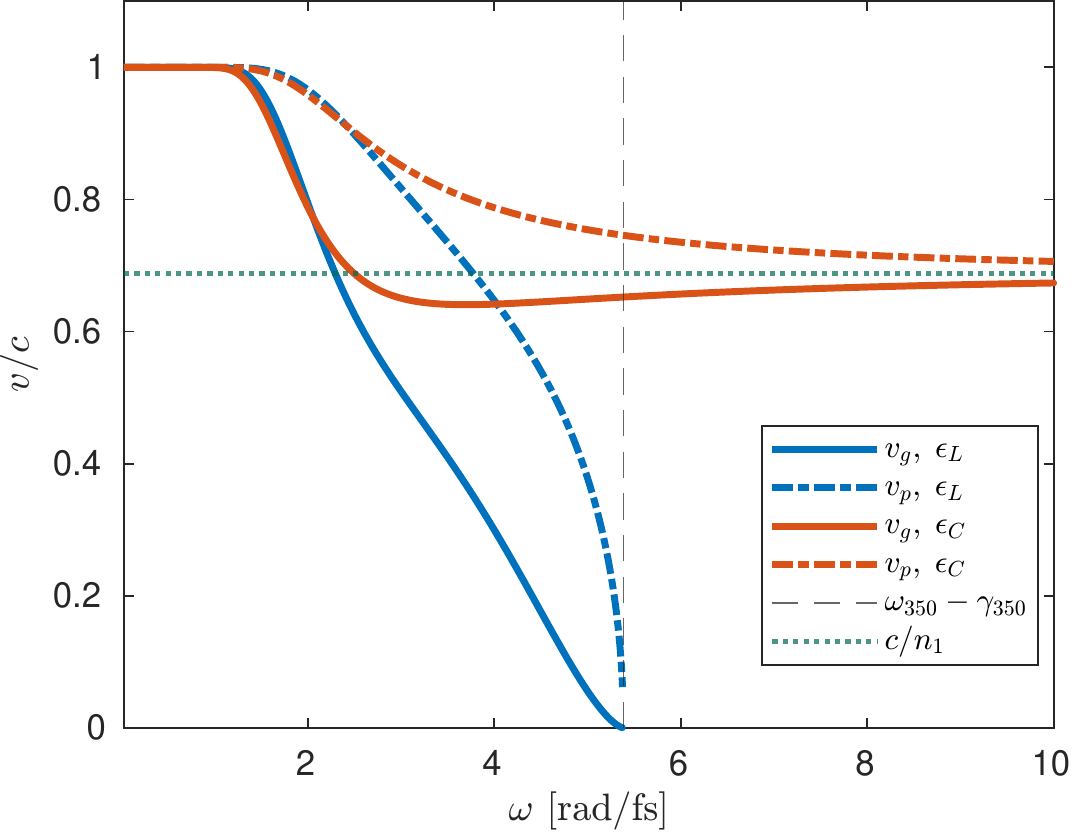}
    \label{Fig:Vel}}
  \caption{Dispersion relation (a) and group and phase velocities (b)
    of the ONF's fundamental mode as a function of the field's
    frequency and using the constant $\epsilon_{C}$ and Drude-Lorentz
    $\epsilon_{L}$ dielectric functions. In Fig.~\ref{Fig:DispRel}
    the solid curves depict the mode's dispersion relation for both
    dielectric functions, while the dashed lines represent their low
    and high frequency asymptotic behaviours corresponding to that of
    vacuum and an infinite dielectric medium with dispersion relation
    $n_{1}(\omega)\omega/c$. In Fig.~\ref{Fig:Vel} the solid and dashed
    lines portray the mode's group and phase velocities normalized
    with respect to $c$, while the dashed vertical and horizontal
    lines correspond to the ONF's resonant frequency and the value of
    the asymptotic velocity when employing the constant dielectric
    function $v_{\infty} = c/n_{1}$, respectively.}
    \label{Fig:DRGV}
\end{figure*}

Since the Hamiltonian under the RWA preserves the total excitation number, we consider the evolution of the following state in the single excitation manifold
\begin{align}
    \ket{\psi(t)} & = \sum_{m = 1}^{2} c_{m}(t) \sigma^{\dagger}_{m} \ket{gg}\otimes \ket{\{0\}} \notag \\
    & \quad \quad + \sum_{\mu} c_{\mu}(t)\ket{gg}\otimes\ket{1_{\mu}},
\end{align}
where $c_{m}$ and $c_{\mu}$ are the atomic and field excitation
probability amplitudes, respectively. Assuming that both atoms are
prepared with their electric dipoles pointing along the radial
direction $\bar{p}_{m} = \bar{p}_{n} = p \hat{r}$ and separated at the
same distance from the surface of the fiber $r_{m} = r_{n} = R$, we
derive the equation of motion for the atomic amplitudes using
Eq.(\ref{Hamiltonian}). Formally integrating for the field excitation
probability amplitudes and substituting them in the equation for
$c_{m}$ we obtain \small
\begin{align}
\dot{c}_{m}(t) & = -\sum_{n = 1}^{2}\int_{0}^{t}dt' F_{mn}(t-t')c_{n}(t'), \label{EvEq} \\
F_{mn}(t) & = \sum_{\mu}G_{\mu m}G_{\mu n}^{*}e^{-i(\omega-\omega_{0})t} \notag \\
& =  \int_{0}^{\infty}d\omega \; e^{-i(\omega-\omega_{0})t}  S(\omega,R)\cos[\beta(\omega)d]\cos(\phi_{m} - \phi_{n}), \label{Integrand}\\
S(\omega,R) & = \Big(\frac{\vert p\vert^{2}}{\pi\epsilon_{0}\hbar}\Big) \omega\frac{\partial \beta}{\partial \omega} \abs{e_{r}(\omega,R)}^{2} \label{spectralDensity}.
\end{align}
Here, $S(\omega,R)$ is the one-point
spectral density of the guided mode. $F_{mn}(t)$ represents the zero temperature correlation function at the position of each atom $(m = n)$ and between the two atomic positions $(m \neq n)$ separated a distance
$d = \abs{z_{m}-z_{n}}$ along the ONF and a distance $R$ from
its surface. The correlation function is given by the Fourier
transform of $S(\omega,R)\cos[\beta(\omega)d]\cos(\phi_{m}-\phi_{n})$. When $m = n$, its real
part is associated with the spontaneous decay of a single atom into
the fundamental mode and its imaginary part to the Lamb shift induced
by the ONF. When $m \neq n$, the correlation function is associated
with the influence one atom excerpts on the other, with its imaginary
part giving rise to the so called dipole-dipole interaction. Since the angular coordinates of the atoms are not coupled to the frequency components of the field, the shape of the two-point correlation function is not affected by the particular choice of this coordinates (only its
strength) and therefore, we set $\cos(\phi_{m}-\phi_{n}) = 1$. When
$F_{mm}(t)$ is a Dirac delta centered at zero and $F_{mn}(t)$ is given by two Dirac deltas displaced by the retarded time, we recover the results in \cite{Sinha2020}.

\section{Results}
\label{section:Results}

\begin{figure*}[t]
    \centering
    \subfloat[]{\includegraphics[width = 3 in]{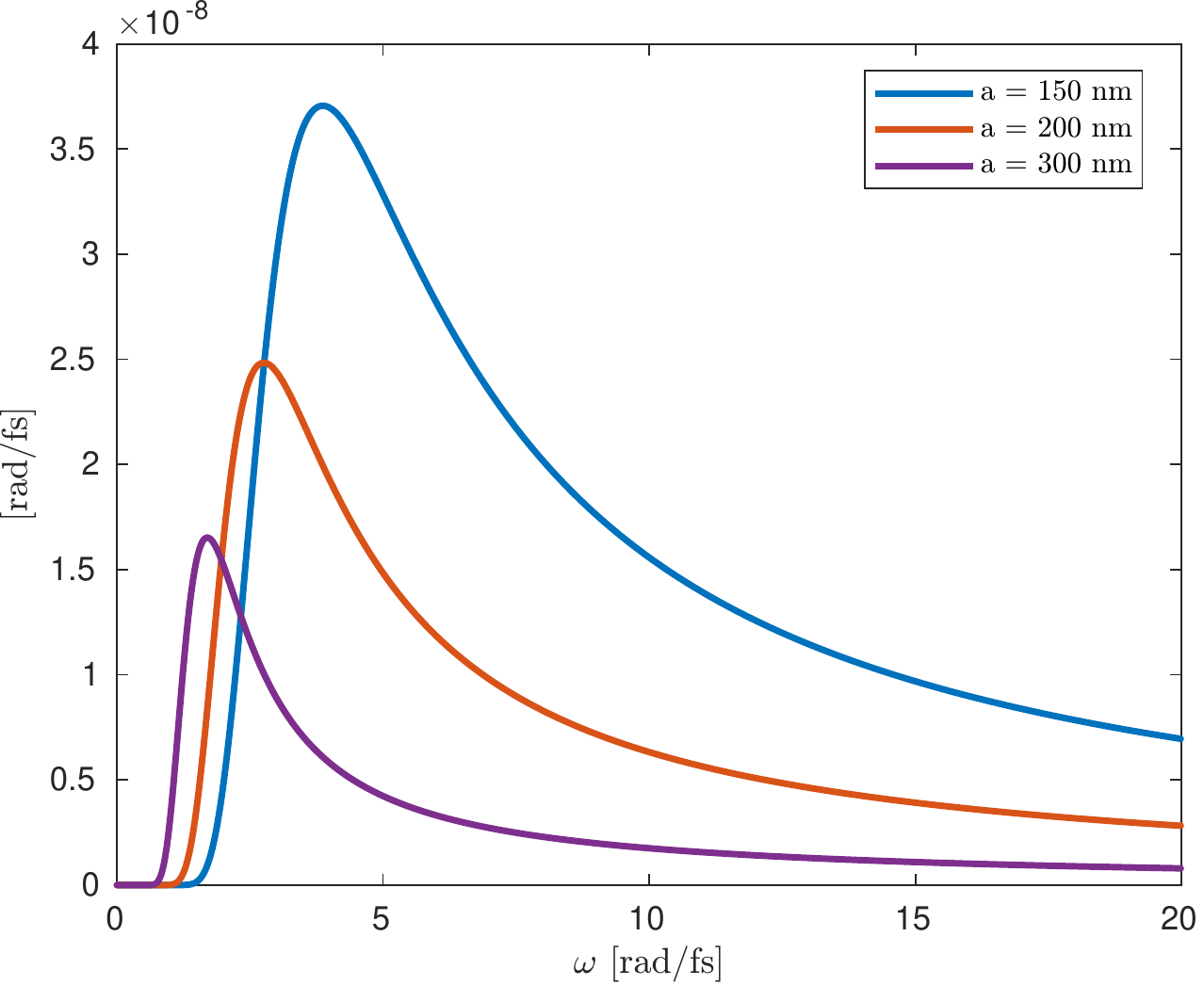}
    \label{Fig:SDN}}
    \subfloat[]{\includegraphics[width = 3 in]{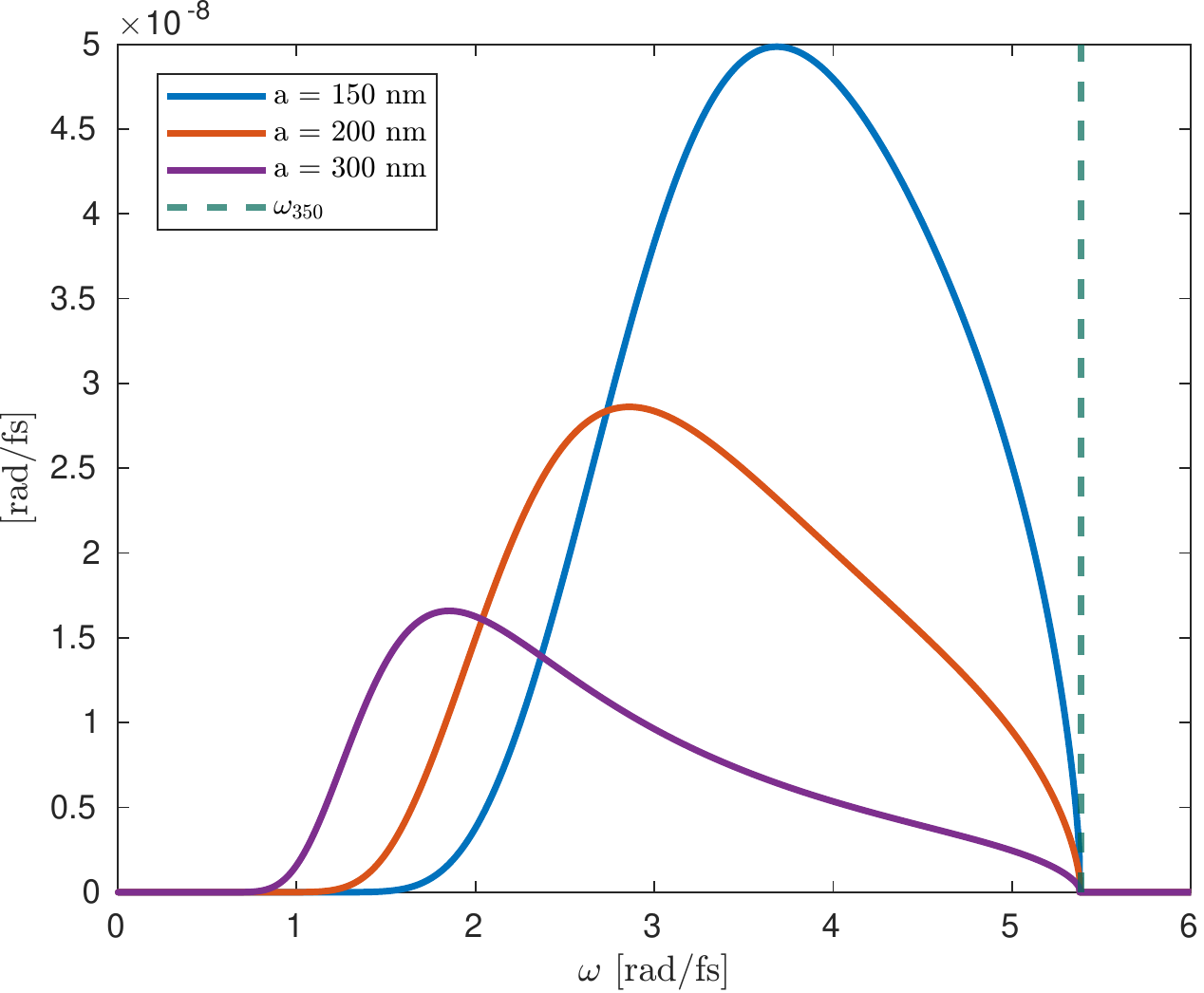}
    \label{Fig:SDL}}
  \caption{One-point spectral density $S(\omega,R = 100 \; \text{nm})$
  for the constant $\epsilon_{C}$ (a) and the DL $\epsilon_{L}$ (b) dielectric functions as a function of the ONF's radius $a$.}
    \label{Fig:SD}
\end{figure*}

\subsection*{Dispersion relations, spectral density and correlation functions}

In order to calculate the correlation functions $F_{mn}(t)$, we first need to solve Eq.~(\ref{EVE}) to obtain the dispersion relation $\beta(\omega)$, shown in Fig.~\ref{Fig:DispRel}. Since the fraction of the guided electromagnetic field contained inside the dielectric nanofiber is inversely proportional to its wavelength, the dispersion relation of the guided mode is asymptotically bounded between the vacuum one at low frequencies and that of a pure dielectric medium, $n_{1}(\omega)\omega/c$, at large frequencies. Fig.~\ref{Fig:Vel} shows the group $v_{g} = (\partial \beta/ \partial \omega)^{-1}$ and phase
$v_{p} = \omega/\beta$ velocities for both dielectric functions we consider. When using the DL model, both group and phase velocities becomes negligible as the frequency approaches the resonance frequency of the
ONF. For frequencies
greater than $\omega_{350}$, the anomalous dispersion phenomenon breaks
down the guiding condition of the ONF, $n_{1}(\omega)-n_{2}>0$. On the other hand, when considering a constant dielectric function, the propagation constant $\beta(\omega)$ increases linearly with frequency and both group and phase velocity coincide, asymptotically approaching the speed of light inside the dielectric, $c/n_{1}$.

Using the numerically calculated $\beta(\omega)$ and
Eq.~(\ref{spectralDensity}) we plot the one-point
spectral density $S(\omega,R)$ in Fig.~\ref{Fig:SD}. For $T=0$ the spectral density is the inverse
Fourier transform of the correlation function
\cite{RevModPhys.89.015001}, which implies that for constant dielectric
function (see Fig.~\ref{Fig:SDN}) large frequency contributions lead to correlation functions localized in time. On the other hand, the time-domain correlations in the DL model are much broader since the spectral density cuts off at the resonant frequency of the ONF (see Fig.~\ref{Fig:SDL}). 

We discretize the integration over $\omega$, allowing us to compute $F_{mn}(t)$ with a fast Fourier transform in an equally spaced time grid $\{t_{j} \vert \; t_{j} = t_{1} + (j-1) \Delta t, \; j \in \mathbb{N} \}$. 
The fact that the propagation constant $\beta(\omega)$ becomes large near the dielectric resonance in the DL model complicates the calculation of the two-atoms correlation function. The $\cos[d\beta(\omega)]$ factor in the two-point spectral density in Eq.~(\ref{Integrand}) oscillates increasingly fast as the integration approaches the resonant frequency, resulting in its
sampling above Nyquist frequency computationally expensive. In order to solve this, preventing the phenomenon of aliasing and considering that our field theory is incapable of describing the effects of high absorption and dispersion, we introduce a hard cutoff far
below $\omega_{350}$ at one of the zeros of the $\cos[d\beta(\omega)]$
factor, which is chosen such that the variation in the results obtained with higher frequency zeros is negligible \cite{deVega17}. We
note that, in contrast to the free space case \cite{carmichael1999statistical,PhysRevA.30.568}, the spectral densities obtained from the guided modes of the ONF do not diverge at high frequencies, producing correlation functions with a finite width even at zero temperature.

\begin{figure*}[t]
  \centering \subfloat[]{\includegraphics[width = 3 in]{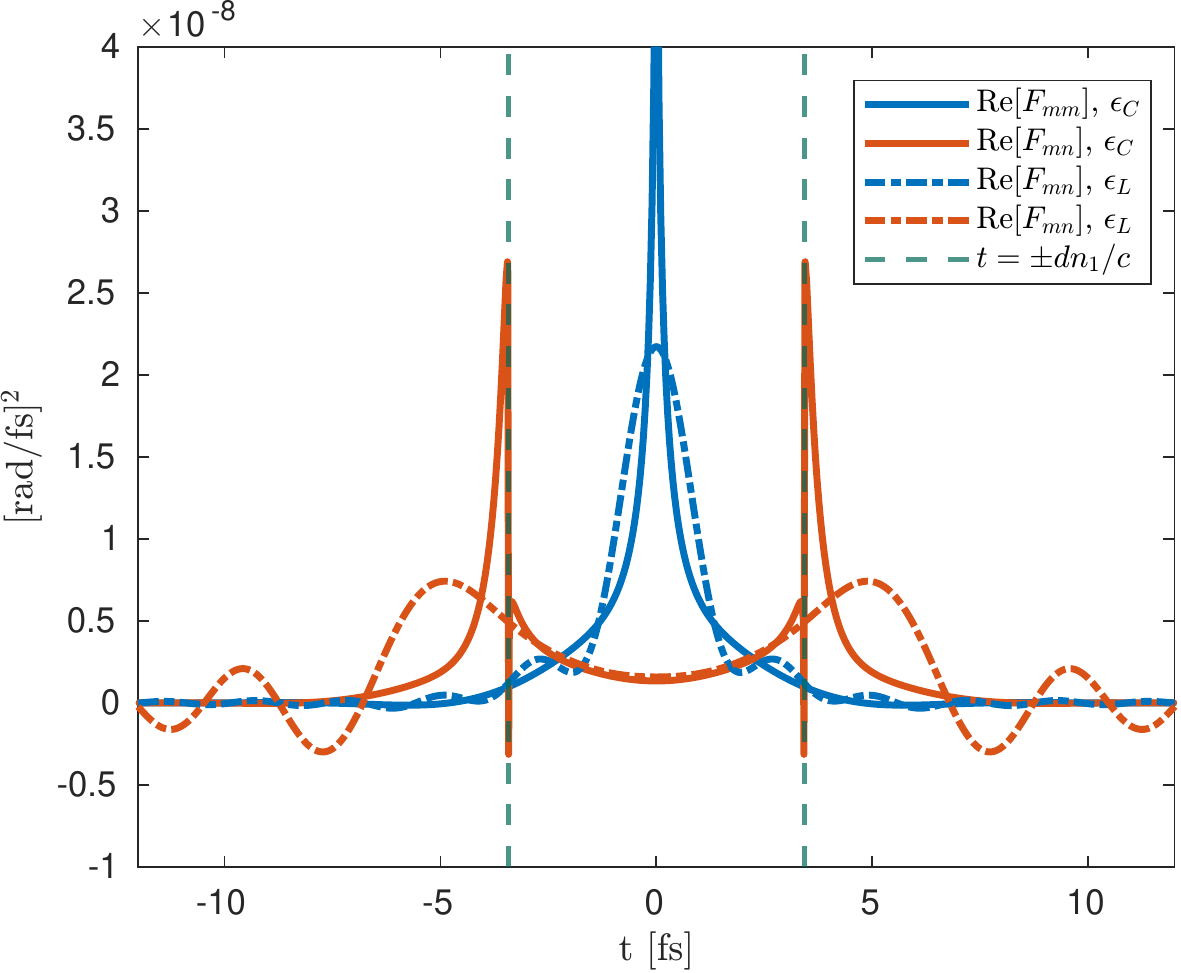}
    \label{Fig:RCF}} \subfloat[]{\includegraphics[width = 3
    in]{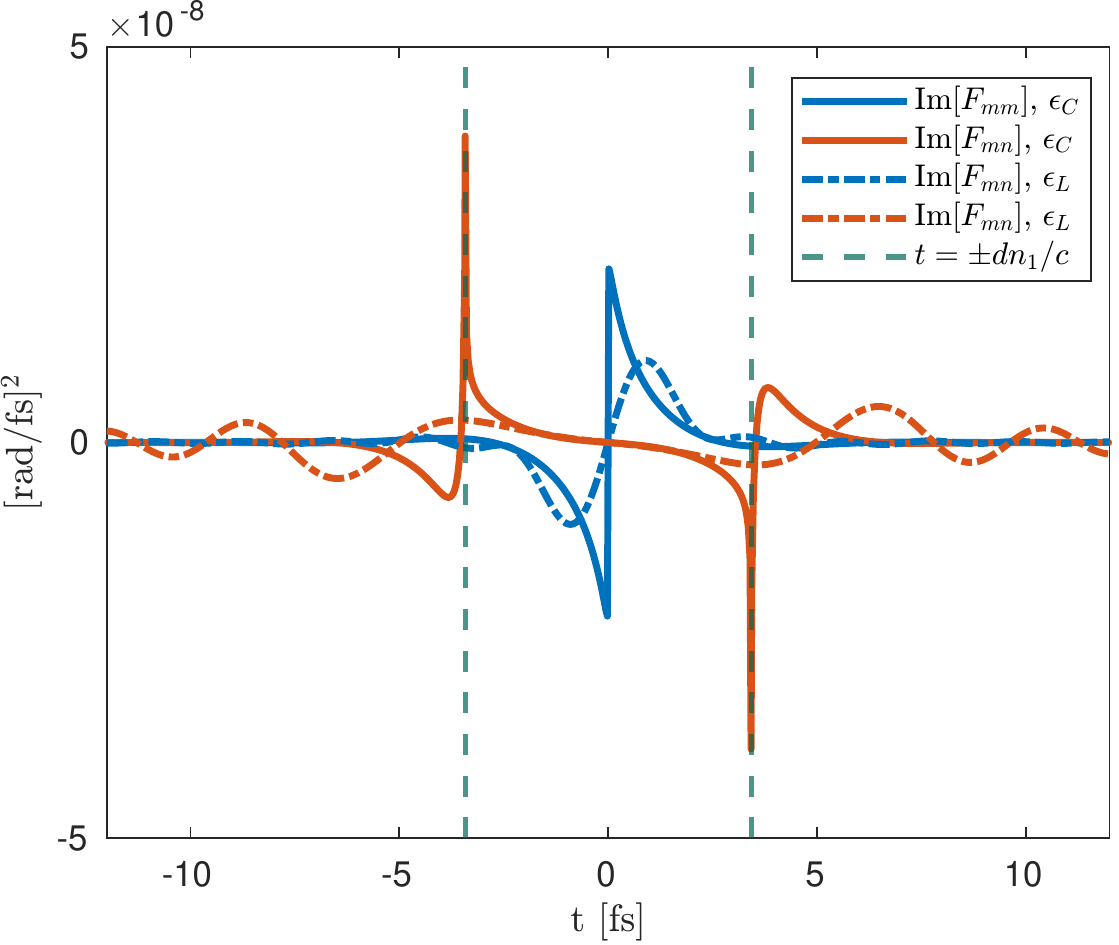}
    \label{Fig:ICF}}
  \caption{ Real (a) and imaginary (b) parts of the fundamental mode's
    correlation functions $F_{mn}(t)$ for a 200 nm-radius ONF in the
    presence of one ($m = n$) and two ($m\neq n$) atoms, both
    separated 100 nm away from the fiber's surface and a distance
    betweem them of an atomic resonant wavelength
    $d = 2\pi/\beta_{0} = 780 \; \text{nm}$ along the fiber's axis.
    Here, the solid curves depict the correlation functions computed
    using the constant dielectric function $\epsilon_{C}$, while the
    dashed ones portray those obtained with the DL function
    $\epsilon_{L}$. Note that for a costant dielectric function the
    correlation functions have peaks with small widths, resembling
    Dirac delta functions. For the DL dielectric function the two-atom
  correlation function has decaying  oscillations that is not clear
  how to approximate with Dirac delta functions.}
    \label{Fig:CF}
\end{figure*}

Figure~\ref{Fig:CF} shows the real and imaginary parts of the correlation function of the fundamental mode for one and two atoms 100 nm away from the fibre surface separated at the resonant atomic wavelength $d = 780$ nm, for a fiber radius of 200 nm. When using a constant dielectric
function we obtain a one atom correlation functions with a sharp and well localized peak at the origin and a two atoms correlation function
with two peaks separated by a time difference $t = 2 d n_1/c$.
% Note that in \cite{Sinha2020} the distance between the peaks in the
% correlation function is given by $t = 2 d/v_{g}$, which coincides with
% the time where collective effects arise. In that sense, our results
% suggest the communication between the atoms via the fundamental guided
% mode of the ONF occurs at rates different from those given by the
% mode's group and phase velocities.
% We will show in the next section that the group velocity is the
% correct choice.

For a constant dispersion relation, the width of the correlation function, which is a measure of the correlation time of the field and one of the sources of non-Markovian effects in the atomic dynamics, is less than
$0.05\;\text{fs}$. This implies that, in the case of one atom, the Markov approximation is justified at zero temperature, a result whose foundation in its vacuum counterpart has been widely discussed by many
authors such as Carmichael \cite{carmichael1999statistical} and only demonstrated recently by Rivas et al. \cite{Rivas_2010}. For two
atoms, the correlation function has two narrow peaks resembling two Dirac deltas. However, the time interval between them, given by $t = 2 d n_1/c$, differs from the intuitive assumption of atom-atom communication at a group velocity, leading to $t = 2 d/v_{g}$. We further investigate this in the next
section. 

When using the DL dispersion relation the resulting correlations present an oscillatory behaviour resembling $\text{sinc}(t)$ functions whose width is approximately
$1.5\; \text{fs}$ in both the one and two atoms situations. This validates the Markovian approximation in for a single atom. Contrary to the case with a constant dielectric function, it's not clear how to associate a communication time between
the atoms. In the light of these results, we integrate Eqs.(\ref{EvEq}) to appreciate the influence of the correlation functions on the collective atomic evolution.

\begin{figure*}[t]
    \centering
    \subfloat[]{\includegraphics[width = 3.07 in]{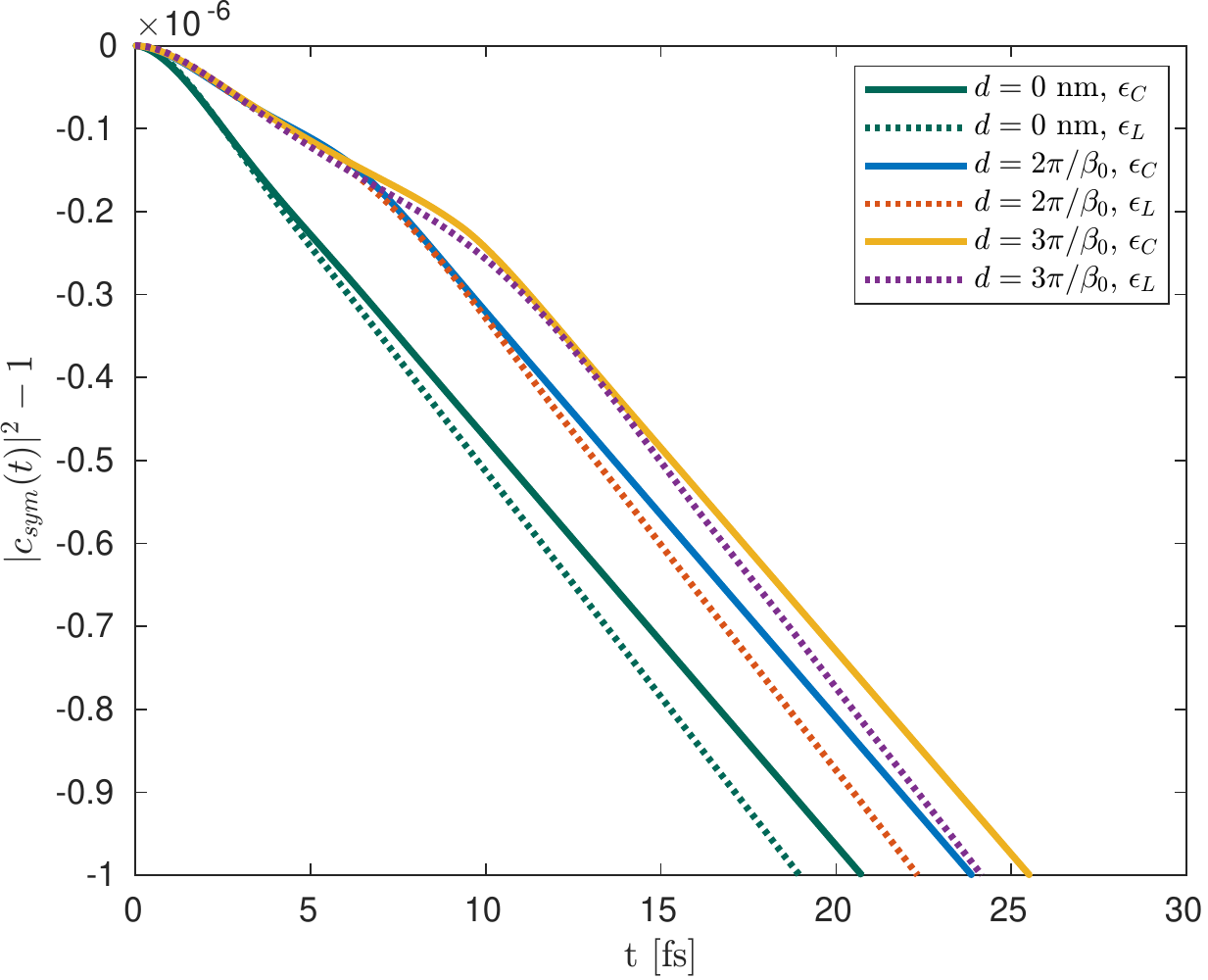}
    \label{Fig:SupPop}}
    \subfloat[]{\includegraphics[width = 3 in]{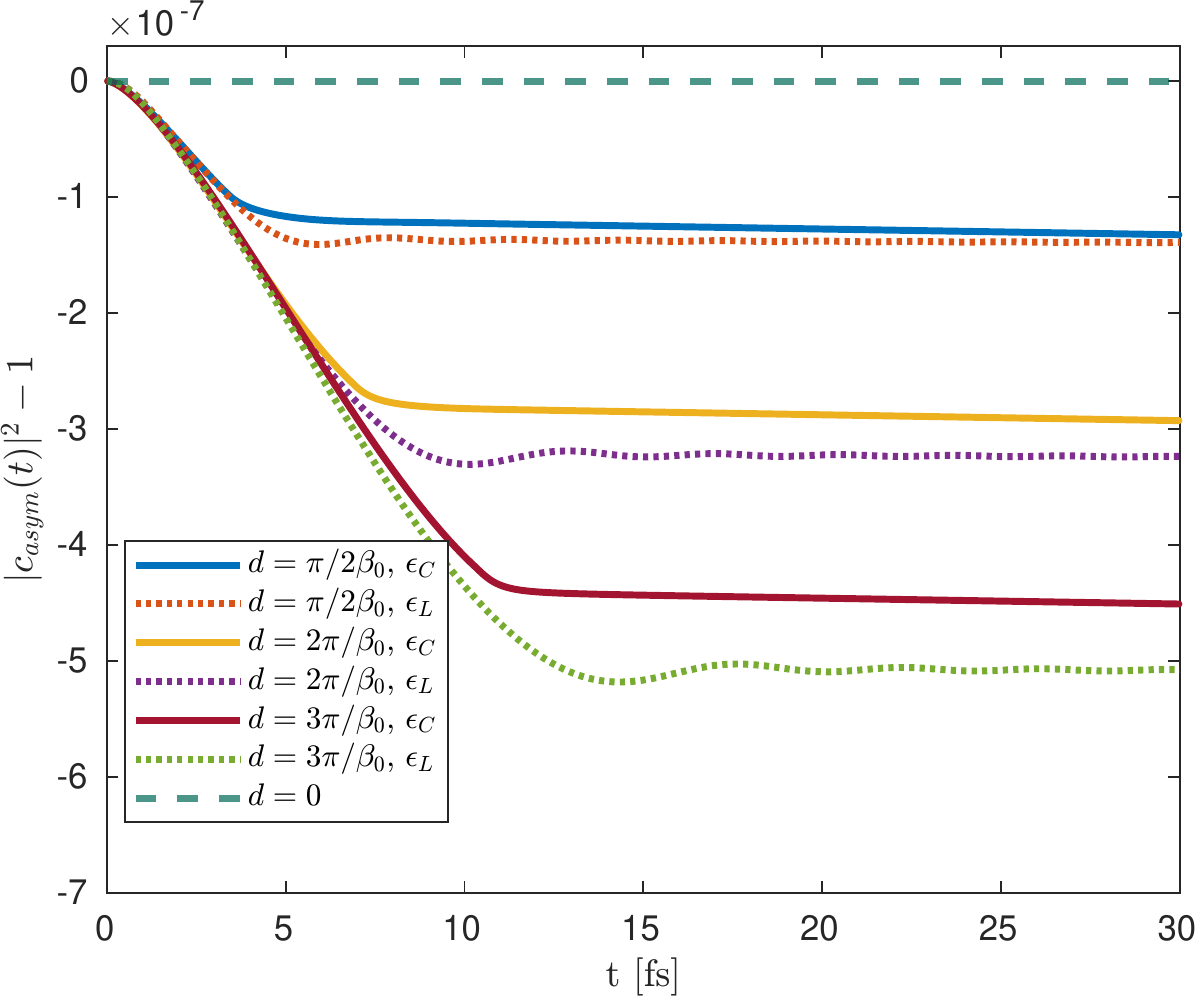}
    \label{Fig:SubPop}}\\
    \subfloat[]{\includegraphics[width = 3 in]{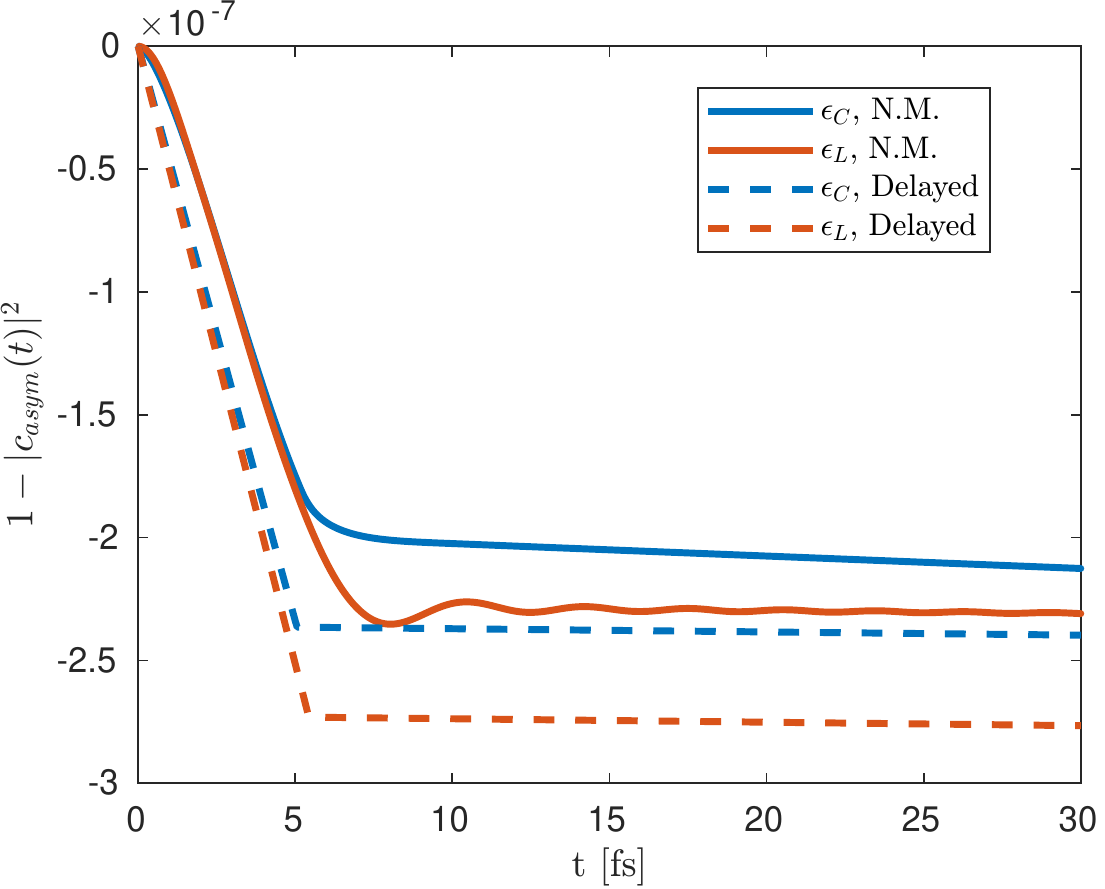}
    \label{Fig:EvEx}}
  \caption{Collective excitation probabilities for initial
    symmetric (a) and antisymmetric (b) states for atoms 100 nm away
    from a 200 nm-radius ONF for several equilibrium separations
    $d_{\text{eq}} = n\pi/\beta_{0}$. In both figures, the solid
    curves refer to the emission probabilities using the constant
    dielectric function, while the dashed ones depict those obtained
    with the DL dielectric function. Note that the transition from
    independent to collective decay is smooth, taking a longer time
    for the DL dielectric function. In (c) we compare the collective
    excitation probabilities calculated with different correlation
    functions for an antisymmetric state for atoms separated at
    $3\pi/\beta_{0}$ using both dielectric functions. The solid curves
    result from ONF's correlation functions, while the dashed ones
    were obtained using displaced Dirac delta correlations. Using
    Dirac delta correlations we obtain an instantaneus establishment
    of a not decaying collective behavior. For the other two
    correlation functions the collective behavior continues to decay,
    although slowly. }
    \label{Fig:PopEv}
\end{figure*}

\subsection*{Atomic dynamics}
Using the numerically calculated field correlation functions, we solve for the dynamics of two atoms initialized in the symmetric or antisymmetric states
$\ket{\psi_{\pm}} = (\ket{eg} \pm
\ket{ge})/\sqrt{2}$. We apply the trapezoidal rule twice on the right hand side of Eq.~(\ref{EvEq}) to numerical solve for the evolution of the atomic excitation probability amplitudes (see appendix \ref{appendix:solutioneq} for details). For simplicity, we analyze our results neglecting the atomic decay into other modes to better present the phenomenology without losing generality.

\begin{figure*}[t]
    \centering
    \subfloat[]{\includegraphics[width = 3 in]{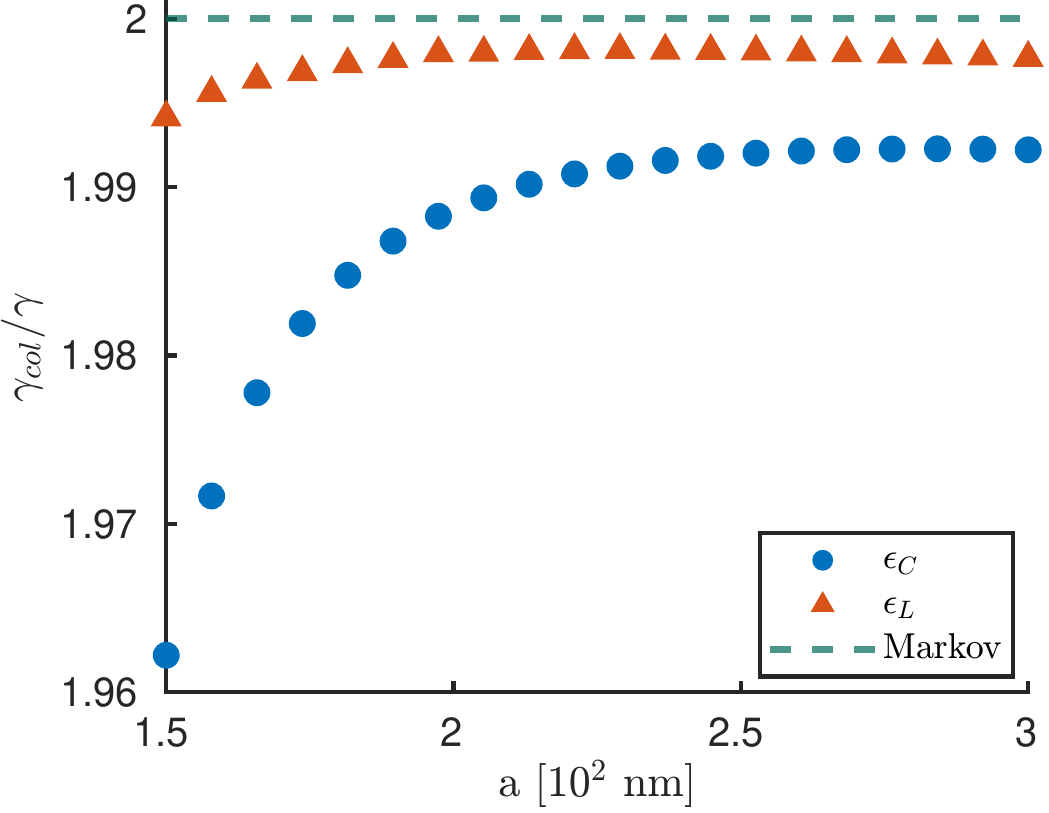}
    \label{Fig:CocSup}}
    \subfloat[]{\includegraphics[width = 3 in]{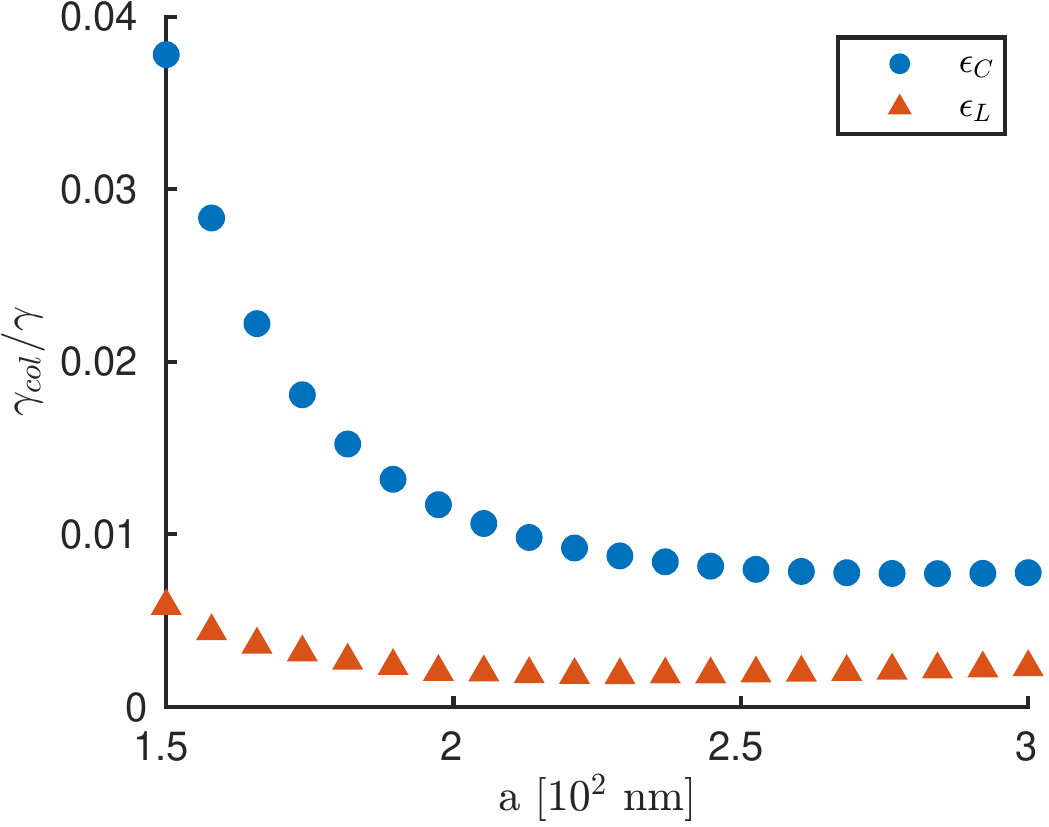}
    \label{Fig:CocSub}}
    \caption{Collective decay rates of superradiant (a) and subradiant (b) atoms (normalized with respect to the decay rate of an independent atom in the vicinity of the ONF, $\gamma$) separated at the resonant atomic wavelength $d = 2\pi/\beta_{0} = 780 \;\text{nm}$ and at 100 nm away from the ONF's surface as a function of the fiber's radius. In both figures, the circular dots depict the collective rates using the constant function, while the triangular portray the results obtained with the DL dielectric function.}
    \label{Fig:CocCol}
\end{figure*}

As Fig.~\ref{Fig:PopEv} shows, the collective excitation probabilities
$\vert c_{\pm}(t)\vert^{2}$ behave as that of
independent emitters before the communication
between the atoms is established. After that, collective features manifest as decay rates close to twice and zero times the natural decay rate for the symmetric and antisymmetric
states respectively, i.e. super and subradiance.
The superradiant decays in Fig.~\ref{Fig:SupPop} display only quantitative differences between both models, while the onset of subradiance inherits the
behavior of the correlation functions used to calculate them, as Fig.~\ref{Fig:SubPop} shows. 
In contrast with the Markovian case, where the collective behavior starts instantaneously upon atom-atom communication, there is a smooth transition from independent to collective decay. Fig.~\ref{Fig:EvEx} shows an example comparing the
two behaviors. Note that, although the atoms are prepared in the subradiant state and at distances which are integer multiples of $\pi/\beta_{0}$, our solutions predict that they must radiate, even if it is at a rate several orders of magnitude below that of a single atom. This contrasts with the picture established in \cite{Sinha2020}, where the emission process is completely inhibited in this situation.

We estimate the modified collective decay rates from the solutions by
fitting the data to a straight line $P(t) = \gamma t + P_{0}$ at times
greater than 300 fs, so that we can assure that the collective
behavior has been fully established. We show in Fig.(\ref{Fig:CocCol})
the quotient between the symmetric and antisymmetric collective decay
rates and that of a single atom as a function of the ONF's radius. We
find that the quotients are independent from the separation between
the atoms and that the deviations from what is obtained with the
Markovian approximation become apparent for radii approximately less
than 4 times the resonant wavelength of the atomic transition, with
variations up to 0.5\% and 4\% for the DL and constant dielectric functions, respectively. In spite of the minor differences between our results and the predictions given by the Markovian approximation, similar differences of a few percent of the decay rate have been measured for single atoms around an ONF \cite{Solano2019}. However, collective effects require a precise positioning of the atoms, hindering its observation for atoms along a nanofiber, but feasible in other wQED platforms. For ONF radii smaller than 150nm, the calculations were not carried out because of the increasing difficulty involved in computing the dispersion relation of the field.

\begin{figure*}[t]
    \centering
    \subfloat[]{\includegraphics[width = 3 in]{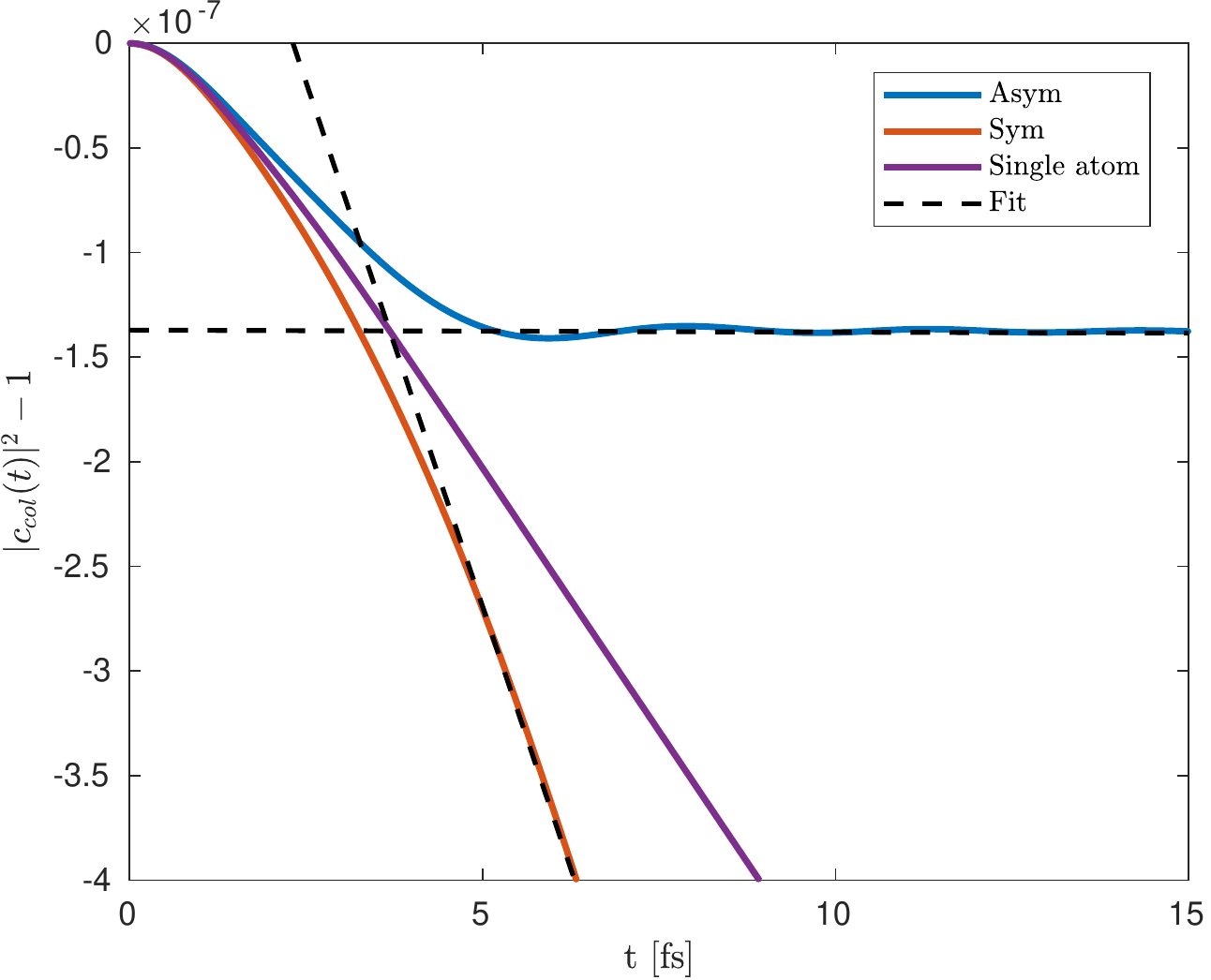}
    \label{Fig:VcomEx}}
    \subfloat[]{\includegraphics[width = 3 in]{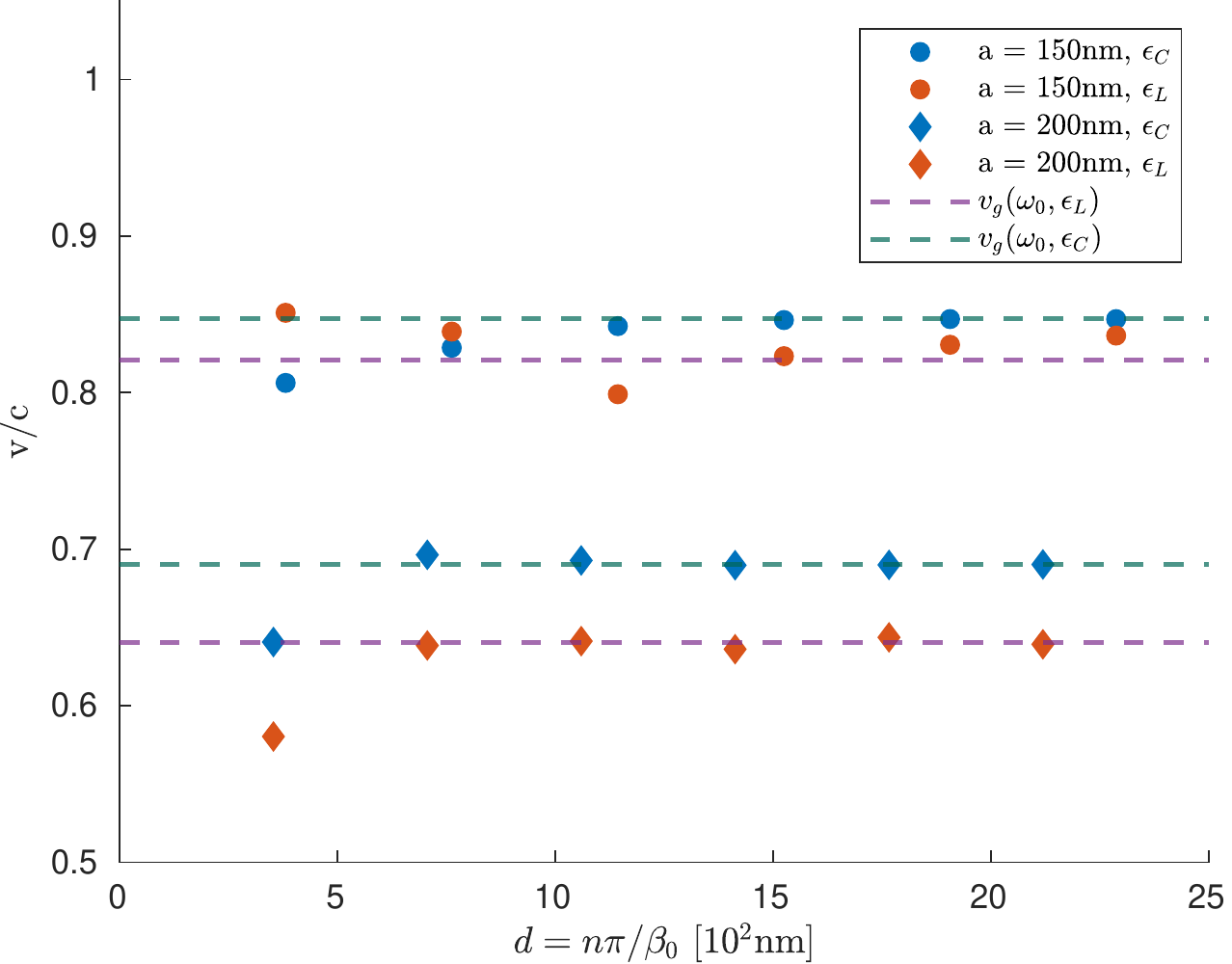}
    \label{Fig:VcomRes}}
    \caption{ (a) Intersection of the linearized collective emission probabilities for initially symmetric and antysimmetric atomic states with the single atom emission probability for atoms separated at their resonant wavelength $d = 2\pi/\beta_{0}$ at 100 nm away from the surface of a 200 nm radius ONF modeled with the DL dielectric function $\epsilon_{L}$. (b) Atom-atom communication speed for atoms separated at
       integer multiples of $\pi/\beta_{0}$ at 100 nm away from the
       ONF's surface. The circular dots and rhombuses portray the
      results for a 150 nm and a 200 nm fiber's radius, respectively.
       The dashed horizontal lines are the group velocities for the
       atom's resonant frequency and a given dielectric function. The
       speeds are normalized with respect to the speed of light in
       vacuum.}
    \label{Fig:VelCom}
\end{figure*}

%\begin{figure}[h!]
%\centering
%     \includegraphics[width = 3 in]{VelCom.pdf}
%     \caption{Atom-atom communication speed for atoms separated at
%       integer multiples of $\pi/\beta_{0}$ at 100 nm away from the
%       ONF's surface. The circular dots and rhombuses portray the
 %      results for a 150 nm and a 200 nm fiber's radius, respectively.
  %     The dashed vertical lines are the group velocities for the
   %    atom's resonant frequency and a given dielectric function. The
    %   speeds are normalized with respect to the speed of light in
    %   vacuum.} \label{Fig:VelCom}
%\end{figure}

\begin{figure}[t]
\centering
     \includegraphics[width = 3 in]{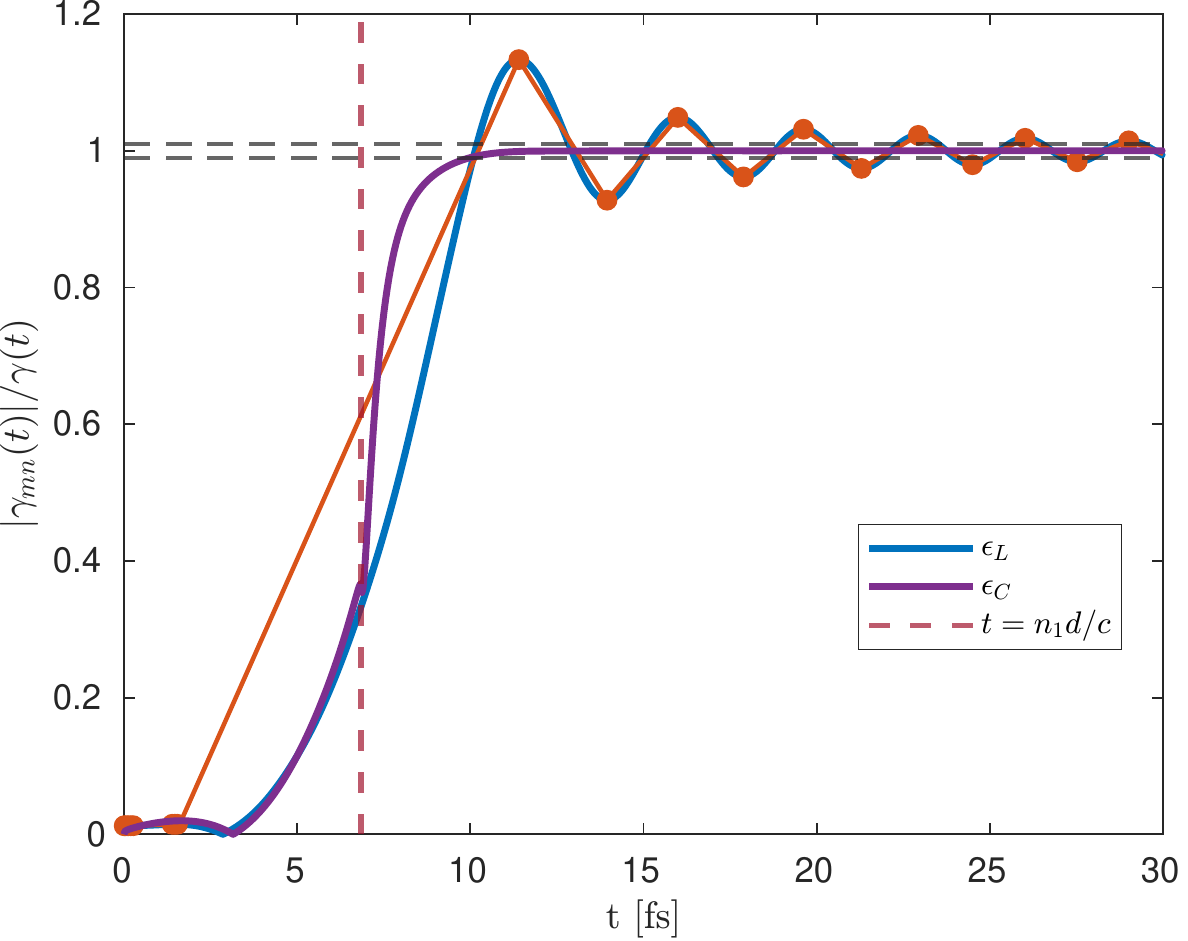}
     \caption{Absolute value of the spontaneous decay rate of an atom due to the influence of the second $\gamma_{mn}(t) = \int_{0}^{t}dt' \Re[F_{mn}(t-t')]$ (normalized with respect the decay rate of an independent atom in the vicinity of the ONF, $\gamma(t)$) for atoms separated at twice the atomic resonant wavelength $d = 4\pi/\beta_{0}$ at 100 nm away from the surface a 200 nm ONF for both dielectric functions. The dots represent the local maxima and minima of the curve obtained with the DL function and the time of establishment of the collective regime $t_{\text{Est}}$ in this case is defined as that at which the middle point of the line connecting two successive maximum and minimum lies within 0.01 from $\abs{\gamma_{mn}(t)}/\gamma(t) = 1$. For the constant function, we consider the establishment time to be when the quotient reaches 0.99.}
     \label{Fig:EstEx}
\end{figure}

\begin{figure*}[t]
    \centering
    \subfloat[]{\includegraphics[width = 3 in]{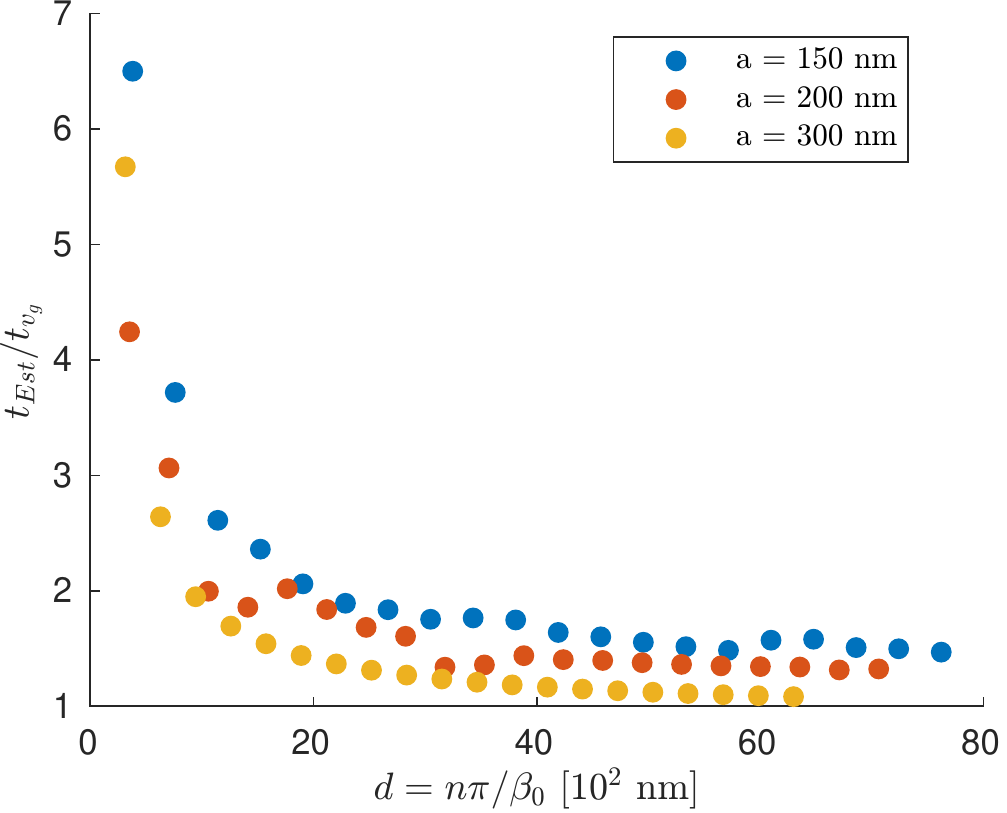}
    \label{Fig:EstablishN}}
    \subfloat[]{\includegraphics[width = 3 in]{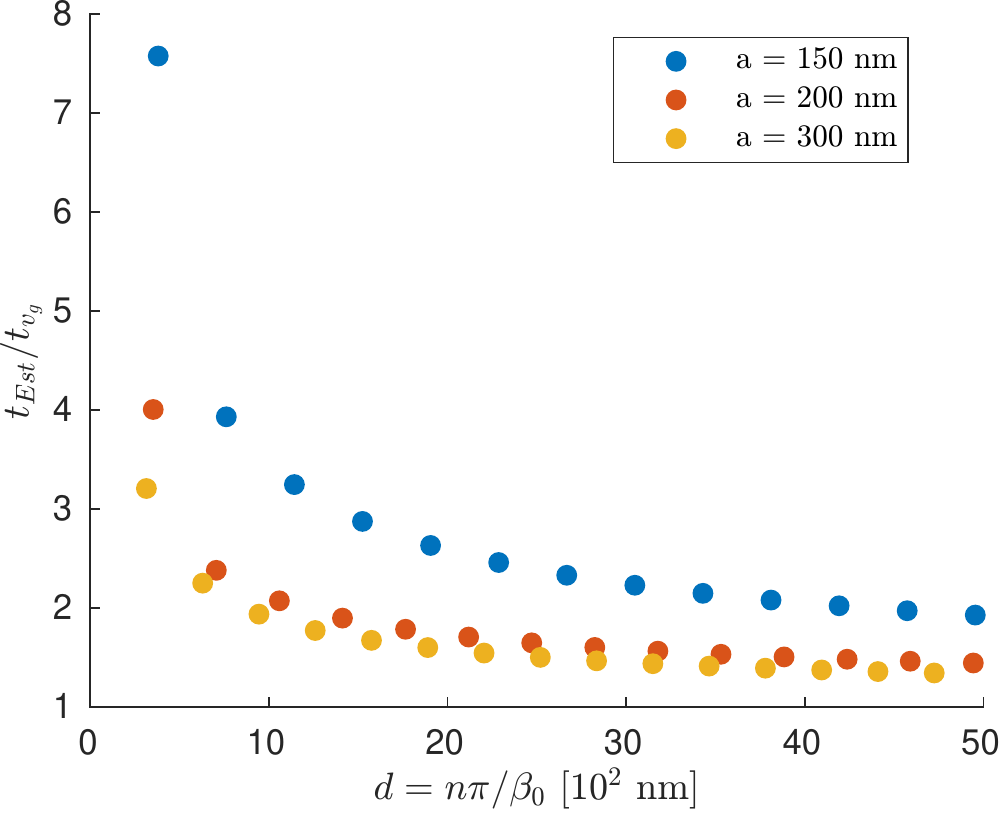}
    \label{Fig:EstablishL}}
    \caption{Times of establishment of the collective decay $t_{\text{Est}}$ for the constant $\epsilon_{C}$ (a) and the DL $\epsilon_{L}$ (b) dielectric functions. The atoms are separated at integer multiples of $\pi/\beta_{0}$ and at 100 nm away from the fiber's surface. The results are normalized with respect to the travel time of a photon moving between the atoms at the group velocity $t_{v_{g}} = d/v_{g}(\omega_{0})$.}
    \label{Fig:Establishment}
\end{figure*}

\subsection*{Onset of the collective decay}

Figure~\ref{Fig:PopEv} shows that it is unclear when the
collective behaviour is fully established. In Ref.~\cite{Sinha2020} the transition from independent emitters to the collective regimes
occurs instantaneously at the time $t = d/v_{g}(\omega_{0})$, which is
the time displacement between the Dirac deltas in the correlation functions, commonly introduced ad hoc. To estimate a time where the collective behavior starts, we extrapolate to earlier times the exponential decay observed at times beyond 300 fs and define the communication time $t_{\text{com}}$ to be that at which the single atom and the approximated long-time collective emission probabilities intersect.
Fig.~\ref{Fig:VcomEx} shows an example of this procedure and
Fig.~\ref{Fig:VcomRes} portrays the communication rate
$v_{\text{com}} = d/t_{\text{com}}$ as a function of the separation
between atoms, the ONF radius and both dielectric functions. We
find that $v_{\text{com}}$ is independent of the initial state of the atoms and that deviations from the group velocity are significant only for separations less than the atomic resonant wavelength
$\lambda_{0} = 2\pi/\beta_{0}$, the exception being the rates
calculated with the DL function for a 150 nm ONF, which are
attributable to the significant oscillations in its correlation
functions. This offers a satisfactory justification for approximating
the correlation function with two Dirac delta peaks separated by the
time the field propagates between the atoms at group velocity even if
it is the phase velocity that appears in the Hamiltonian through its relation to the
propagation constant $\beta(\omega)$.

To estimate the time to establish the collective behavior
$t_{\text{Est}}$, we account for the fact that the
solutions vary insignificantly in magnitude during the transition
from the independent to the collective behavior and thus, we extract
the probability amplitudes in Eq.(\ref{EvEq}) out of the time integral
and study the behavior of the correlation functions integrals. Considering this and the fact that the imaginary parts of the correlation approach their asymptotic behavior in the same time scales as their real counterparts, the evolution
equations become
\begin{align}
    \dot{c}_{m}(t) & = - \big[ \gamma_{mm}(t)c_{m}(t) + \gamma_{mn}(t)c_{n}(t)  \big],\\
    \gamma_{mn}(t) & = \int_{0}^{t}dt' \Re[F_{mn}(t-t')].
\end{align}
Here, in correspondence to the results obtained with the Markovian approximation \cite{2005LeKien}, the collective decay rates of (anti)symmetric states are given by
\begin{align}
    \gamma_{\mp}(t) & = \gamma(t) \mp \abs{\gamma_{mn}(t)}, \label{ColGam}
\end{align}
which in case of the Markovian approximation and separations which are
integer multiples of $\pi/\beta_{0}$, $\abs{\gamma_{mn}} = \gamma$,
and thus, these states correspond to a (sub)superradiant emission.
Fig.~\ref{Fig:EstEx} depicts the time dependent quotient of the decay
rates $\vert \gamma_{mn}(t)\vert /\gamma(t)$ for atoms separated at
twice their resonant wavelength and 100 nm away from the surface of a
200 nm radius ONF for both dielectric functions. When using the
constant dielectric function, the quotient becomes 1 for times greater
than $t = d n_1/c$ and the time of establishment of the
collective behavior is regarded as the time the quotient reaches to
0.99. Meanwhile, for the the DL function, the quotients calculated for
different separations grow to a maximum and converge asymptotically
towards 1 as $t \rightarrow \infty$ while oscillating. In this case, we define the time to establish the collective behavior as the time at which the midpoint of the line
joining two successive maximum and minimum of the curve lies within a
range less than 0.01 from $\vert \gamma_{mn}(t)\vert /\gamma(t) = 1$,
as shown in Fig.~\ref{Fig:EstEx}. Fig.~\ref{Fig:Establishment} shows
the time to establish the collective decay for both dielectric
functions normalized with respect to the time a photon propagates
between the atoms moving at the group velocity, which we denote as
$t_{v_{g}} = d/v_{g}(\omega_{0})$. The results are shown as a function
of the separation between the atoms and the ONF's radius. Again, we
find the time to establish collective behaviours is independent of the initial atomic state and that it decreases to values less than $2t_{v_{g}}$ when
employing the DL function, and almost to $t_{v_{g}}$ when considering the constant dielectric function. Thus, when the atoms are separated
several times their resonant wavelength, the distinction between the independent and collective regimes is established in a time smaller
than $2t_{v_{g}}$.

\section{Conclusions}
\label{section:Conclusions}
In this work, we calculated the correlation functions of the electromagnetic environment provided by the fundamental guided mode of an optical nanofiber. We studied their effects on the collective dynamics of two separated two-level atoms. The width and central position of the correlation functions depends strongly on the
dispersion relation of the waveguide. The correlation functions resemble nascent delta distributions when considering a constant dielectric function. Still, the time difference between their peaks does not coincide with the time that would take for the field to propagate between the two atoms propagating at group or phase velocity. Nevertheless, when studying the dynamics of the atoms, we found that approximating the correlation function with two Delta dirac functions separated by the time the field propagates between the atoms at group velocity is a good approximation, provided the atoms are placed far enough. When the atoms are just a few wavelength apart the intuition from a well defined traveling wavepackage breaks down and it becomes hard to define a unique characteristic timescale to establish collective behaviours. We obtained the collective excitation probabilities of super and subradiant atoms by solving their Schr\"odinger equation in the non-Markovian regime and found that the collective decay rates can differ by less than 5$\%$ and 1$\%$ compared to the Markovian approximation when considering a constant and Drude-Lorentz dielectric functions, respectively. We conclude that the Markov approximation is good enough for state-of-the-art experiments involving atoms around optical nanofibers. However, its validity must be examined in other waveguide QED platforms, considering their particular dielectric function, dispersion relation, and the level of precision the experiments might require.

% Finally, we point out that our results challenge the validity of
% applying the flat spectral density approximation in waveguide QED
% setups when multiple atoms couple with the waveguide, since it assumes
% that only the atomic resonant frequency contributes significantly in
% the coupling between the atoms and the reservoir
% $G_{\mu m}(\omega) \approx G_{\mu m}(\omega_{0})$, leading to delta
% correlations displaced at times given by the phase velocity of the
% field and perfect super and subradiant emission at integer and
% half-integer multiples of the atomic resonant wavelength. While this
% approximation is well suited for vacuum configurations, it is clear
% from our results that careful consideration is required in this
% context if a consistent implementation of both retardation effects and
% a better account of the emission process are expected.

%the relation between the field's dispersion relation and its resulting correlation functions and found that they are essential to

%In order to study the relation between the field's dispersion relation and its resulting correlation functions, we employed two dielectric functions

%This was done using two dielectric functions  

%\begin{figure}[b]
%    \centering
%    \includegraphics[width = 3.4 in]{Alternativa.png}
%    \caption{}
%    \label{Fig:sch}
%\end{figure}

%\begin{figure*}[t]
%    \centering
%   \subfloat[]{\includegraphics[width = 3 in]{Schematic.png}}
%    \subfloat[]{\includegraphics[width = 3 in]{SchematicCircuit.png}}
%    \caption{ }
%    \label{schematic}
% \end{figure*}

%\section{Discussion}

\section*{Acknowledgments} 

We thank K. Sinha for insightful discussions. P.S. is a CIFAR Azrieli
Global Scholar in the Quantum Information Science Program. This work
was supported in part by CONICYT-PAI grant 77190033, FONDECYT grant
N$^{\circ}$ 11200192 from Chile and DGAPA-PAPIIT grant IG101421 from
UNAM, México.

\appendix
\section{Solution of the equations}
\label{appendix:solutioneq}
In order to solve the differential Eq.~(\ref{EvEq}) we transform it in
a linear system of equations, in which the probability amplitudes
evaluated at time grid of the correlation functions are recursively obtained by solving the system \small
\begin{align}
    M \bar{x} & = \bar{y},\\
    \bar{x}^{\intercal} & = [a_{1}^{j+1},a_{2}^{j+1},b_{1}^{j+1},b_{2}^{j+1}],\\
    M & = \mathbf{Id} \notag \\
      & + \Big(\frac{h}{2}\Big)^{2}\begin{bmatrix}
	A(1,1) & -B(1,1) & C(1,1) & -D(1,1) \\
	B(1,1) &  A(1,1) & C(1,1) &  D(1,1) \\
	C(1,1) & -D(1,1) & A(1,1) & -B(1,1) \\
	D(1,1) &  C(1,1) & B(1,1) &  A(1,1)
	\end{bmatrix}, 
\end{align}
 \normalsize where $\mathbf{Id}$ is the $4\times 4$ identity matrix and  \small
\begin{align}
	A(j,k)  & = \Re[F_{mm}(t_{j}-t_{k})], \; B(j,k)  = \Im[F_{mm}(t_{j}-t_{k})],\\
	C(j,k)  & = \Re[F_{mn}(t_{j}-t_{k})], \; D(j,k)  = \Im[F_{mn}(t_{j}-t_{k})],\\
	a_{1}^{j} & = \Re[c_{m}(t_{j})], \;  a_{2}^{j} = \Im[c_{m}(t_{j})],\\
	b_{1}^{j} & = \Re[c_{n}(t_{j})], \;  b_{2}^{j} = \Im[c_{n}(t_{j})],
\end{align} 
\normalsize refer to the real $\Re$ and imaginary $\Im$ parts of both the correlation functions and the atomic probability amplitudes; the vector $\bar{y}$, which takes into account the present and past states of the atomic amplitudes, is defined as
\begin{widetext}
\begin{align}
    \bar{y} & =  \begin{pmatrix}
	[1-Y^{A}]a_{1}^{j} - { [ X_{a_{1}}^{A} - X_{a_{2}}^{B} + X_{b_{1}}^{C} -X_{b_{2}}^{D}] - [Y^{B}a_{2}^{j} -Y^{C}b_{1}^{j}  +Y^{D}b_{2}^{j}] } \\
	[1-Y^{A}]a_{2}^{j} - {[X_{a_{2}}^{A} - X_{a_{1}}^{B} + X_{b_{2}}^{C} -X_{b_{1}}^{D}] - [Y^{B}a_{1}^{j} -Y^{C}b_{2}^{j}  +Y^{D}b_{1}^{j}] } \\
	[1-Y^{A}]b_{1}^{j} - {[X_{b_{1}}^{A} - X_{b_{2}}^{B} + X_{a_{1}}^{C} -X_{a_{2}}^{D}] - [Y^{B}b_{2}^{j} -Y^{C}a_{1}^{j}  +Y^{D}a_{2}^{j}] } \\
	[1-Y^{A}]b_{2}^{j} - {[X_{b_{2}}^{A} - X_{b_{1}}^{B} + X_{a_{2}}^{C} -X_{a_{1}}^{D}] - [Y^{B}b_{1}^{j} -Y^{C}a_{2}^{j}  +Y^{D}a_{1}^{j}] } 
	\end{pmatrix}, \\
    Y^{F} & = \Big(\frac{h}{2}\Big)^{2} [2F(2,1) + F(1,1)],\\
	X_{d}^{F} & = \Big(\frac{h}{2}\Big)^{2} \Bigg\{ 2\sum_{l = 2}^{j-1}[ F(j+1,l) + F(j,l)]d^{l} + [F(j+1,1) + F(j,1)]d^{1}\Bigg\}.
\end{align}
\end{widetext}

\bibliography{ref}

\end{document}